\documentclass[4pt,aps,prx,reprint,superscriptaddress
]{revtex4-2} %

\usepackage[dvipsnames]{xcolor}

\usepackage{dcolumn}
\usepackage{bm}

\usepackage[utf8]{inputenc}
\usepackage[T1]{fontenc}
\usepackage{hyperref}
\usepackage{mathrsfs}
\usepackage{amsfonts}
\usepackage{amsmath}
\usepackage{tcolorbox}
\usepackage{float}

\usepackage{graphicx}
\usepackage{dcolumn}
\usepackage{bm}

\begin{document}


\title{
The specific heat anomaly stems from a third-order phase transition in the 2D lattice sine-Gordon model
}

\author{Loris Di Cairano}              
\email{loris.dicairano@uni.lu}
\affiliation{Department of Physics and Materials Science, University of Luxembourg, L-1511 Luxembourg City, Luxembourg}

\author{Alexandre Tkatchenko}            
\email{alexandre.tkatchenko@uni.lu}
\affiliation{Department of Physics and Materials Science, University of Luxembourg, L-1511 Luxembourg City, Luxembourg}
\date{\today}


\begin{abstract}
The specific heat anomaly (SHA) is broadly observed in statistical mechanics, appearing as a smooth, system-size-independent peak in the specific heat, in contrast to the singular behavior typical of second-order phase transitions (PTs). Its origin remains heavily debated: some attribute it to finite-size effects, others to unidentified phase transitions. Here we investigate SHA using the two-dimensional sine-Gordon (2D-sG) model and microcanonical inflection point analysis (MIPA), uncovering two key results. First, we show that the roughening transition in the 2D-sG model is a genuine third-order PT under MIPA, where the standard thermodynamic quantities remain continuous. This clarifies the ambiguity in the literature, where this transition was often, though inconclusively, attributed to a Berezinskii-Kosterlitz-Thouless (BKT) transition. Through the use of MIPA and a comprehensive analysis of standard thermodynamic observables, we provide a coherent thermodynamic characterization that redefines the nature of this transition.
Second, we find that the SHA is not itself a PT but rather the thermodynamic fingerprint of this third-order transition. These findings clarify the nature of SHA within the 2D-sG model and suggest that similar anomalies in other systems, such as the XY model, may likewise originate from third-order PTs, rather than mere crossovers. Our results provide a consistent thermodynamic interpretation of the SHA and highlight the broader relevance of third-order transitions in systems previously thought to exhibit only low-order or crossover transition.
\end{abstract}

\maketitle

\section{Introduction}

The appearance of a rounded, system-size-independent peak in the specific heat—called the \textit{specific heat anomaly} (SHA) or \textit{specific heat shift}—is a long-standing and debated problem. The SHA is characterized by two peculiar and related features. 

First, despite manifesting a prominent peak, the specific heat remains analytic even in the thermodynamic limit. This behavior is considered anomalous because it cannot be explained by existing theories such as Ehrenfest-Lee-Yang (ELY) \cite{ehrenfest1933phasenumw,lee1952statistical,yang1952statistical} and renormalization group (RG) \cite{kadanoff1966scaling,wilson1975renormalization}, being based on divergences of suitable thermodynamic observables. In fact, in systems undergoing second-order phase transitions (PT), the specific heat always manifests a peak that becomes non-analytical in the thermodynamic limit. In the SHA, even though the size of the system is increased, the peak of the specific heat reaches a value that remains constant, thus preventing identification with a second-order PT. 

Second, the SHA often coexists with a major transition, such as, for instance, the superfluid PT as observed in 2D helium film experiments since the 1960s \cite{brewer1965gapless,van1955susceptibility,gijsman1959magnetic,steele1993precision,nakamura2016possible}. Superfluid transitions are of Berezinskii-Kosterlitz-Thouless (BKT) type \cite{kosterlitz1973ordering,berezinskii1971destruction}, driven by the unbinding of pairs of topological defects (vortex-antivortex pairs) above a critical temperature. Interestingly, the SHA peak often occurs at temperatures above the BKT onset and does seem to have a topological origin. Its emergence has even been interpreted as the signature of a new ``superhexatic'' phase \cite{nakamura2016possible}. Although such interpretations remain controversial \cite{boninsegni2023superfluid,corboz2008phase,ahn2016prediction,moroni2019second,boninsegni2020specific}, it has been suggested that measurements of specific heat alone are insufficient to identify PTs \cite{boninsegni2020specific}. 

Beyond experiments, SHA has also been observed in numerical studies of several 2D lattice models, notably the XY model \cite{tobochnik1979monte,van1981helicity,solla1981vortex,kawabata1982monte} and the 2D sine-Gordon (sG) model \cite{sanchez2000roughening}. Although early studies attributed the SHA in the XY model to finite-size effects \cite{van1981helicity}, later works confirmed its persistence even in very large systems \cite{gupta1988phase,bowen1992monte,ota1992microcanonical,cuccoli1995two,nguyen2021superfluid}. In general, the thermodynamics of the XY model has been deeply studied and the theoretical description of the BKT transition based on RG methods is confirmed by numerical simulations. Moreover, the emergence of the SHA is widely reported and consistent. 

The situation is different for the sG model. In fact, SHA has been detected, but has not been thoroughly studied. Moreover, the transition manifesting in the sG model, called roughness PT, presents evident difficulties in being univocally characterized with the existing theoretical framework. 

The main reason that makes its study particularly compelling lies in the disagreement between the theoretical predictions based on RG methods and the numerical results. This aspect has been noticed and formulated in Ref.~\cite{sanchez2000roughening}: the reliability of RG methods depends on post hoc comparison with simulations, highlighting their limitations in certain contexts. Although the roughness PT in the sG model resembles a BKT-type transition, the predictions based on RG disagree with the numerical results~\cite{falo1991langevin,sanchez1995roughening,sanchez2000roughening}, and a definitive answer to the question regarding the nature of this transition remains elusive.

This ambiguity highlights a broader conceptual tension: the prevailing view associates non-divergent signals in thermodynamic observables with crossover phenomena, implicitly reserving the notion of PT for cases displaying divergences. However, the roughness PT challenges this dichotomy. Despite the absence of divergences, it manifests as a structurally stable, reproducible phenomenon across system sizes. By identifying the roughness PT as a genuine third-order transition through a microcanonical analysis, we show that higher-order transitions must be acknowledged as thermodynamically real, even though they present smooth rather than singular signatures. 

In this work, we propose a thermodynamic answer to this question and a resolution to the SHA problem. To address this puzzle, we apply the \textit{microcanonical inflection point analysis} (MIPA), a system-independent approach based on the microcanonical ensemble, introduced by Bachmann et al. \cite{sitarachu2022evidence,qi2018classification,di2024detecting}. We consider the 2D-sG model as a case study and show how MIPA unifies, under a consistent framework, seemingly separate phenomena, such as the SHA and the roughness PT. 

We report two key findings. First, we show that the roughness transition is a third-order PT. Unlike first- and second-order PTs, this transition lacks non-analyticities, but remains stable with increasing system size. Second, the SHA is not a PT but the thermodynamic signature of this third-order PT. This connection reveals a broader principle: SHA signals higher-order PTs. 

These results suggest that, if the roughness PT is indeed a true PT—as is implicitly accepted—then not all non-divergent signals should be interpreted as crossovers. In particular, while first- and second-order PTs manifest non-analyticities in the thermodynamic limit, true PTs of higher order may be characterized by analytical behavior accompanied by specific, persistent morphological features in entropy derivatives, as exemplified by the roughness PT. Finally, we corroborate our findings by analyzing histograms of sG field configurations and computing scalar observables, independent of MIPA, that confirm its predictions. The generality of this approach allows the results to be extended to other systems where the SHA is observed.

\section{Introduction to the model and microcanonical approach}

The sine-Gordon (sG) field theory discretized on a 2D square lattice is defined as \cite{benfatto2013berezinskii}:
\begin{equation}\label{def:potential_sG}
V_{sG} = \sum_{\bm{n} \in \mathbb{L}} \left[ \sum_{i=1}^2 \frac{(\phi_{\bm{n}+e_i} - \phi_{\bm{n}})^2}{2\pi K} + \frac{g}{\pi} \left(1 - \cos(2\phi_{\bm{n}})\right) \right],
\end{equation}
where \( K \) and \( g \) are positive real parameters, \( \phi_{\bm{n}} \in (-\infty,\infty) \) are scalar fields on a square lattice \( \mathbb{L} \) of size \( N \times N \), and \( \bm{n} + e_i \) denotes a unit shift in the \( i \)-th direction.

The model is of broad interest due to applications in quantum computing \cite{wang2019quantum}, Josephson junction arrays \cite{roy2021quantum,mazo2014sine}, biological systems \cite{wattis2001dynamic,yomosa1983soliton,zhang1987soliton,ivancevic2013sine}, and low-dimensional field theories \cite{cuevas2014sine,mussardo2010statistical,ichinose1994renormalization}. \\
Its thermodynamics has also been studied in the context of surface and crystal growth \cite{safran2018statistical}, where it appears as a modified version of the discrete Gaussian solid-on-solid (DGSOS) model \cite{falo1991langevin}, known to exhibit a roughness (PT) \cite{weeks1979dynamics,chui1976phase}. The DGSOS model is also dual to the XY model \cite{jose1977renormalization}, and by transitivity, the 2D-sG model is expected to undergo a BKT-type roughness transition.\\
Initial studies \cite{falo1991langevin,sanchez1995roughening} confirmed this using the RG method introduced by Chui and Weeks. However, subsequent analysis by Sánchez et al.~\cite{sanchez2000roughening} revealed inconsistencies: RG-based predictions of the transition temperature \( T_R \) using different approaches, such as the Edwards-Wilkinson (EW) equation \cite{edwards1982surface} and the Kosterlitz RG method \cite{kosterlitz1974critical}, yielded conflicting results. \\
The EW-based prediction matched non-perturbative simulations and was supported by the behavior of a (system-dependent) roughness parameter, whereas the Kosterlitz-based estimate did not.\\
The discrepancy was attributed to fundamental differences in how the transition is defined. The Kosterlitz method identifies \( T_R \) via universal scaling of correlation functions, while the EW approach agrees with the high-temperature limit of the mean energy. Although the EW method provides more consistent estimates, it remains perturbative and system-dependent.\\
To overcome these limitations, we apply MIPA, a system-independent method capable of identifying PTs without relying on asymptotic approximations. This motivates our use of the microcanonical ensemble.

\section{Phase transitions in the microcanonical ensemble}
\begin{figure*}[tbh!]%
    \includegraphics[height=10cm, width=18cm,keepaspectratio]{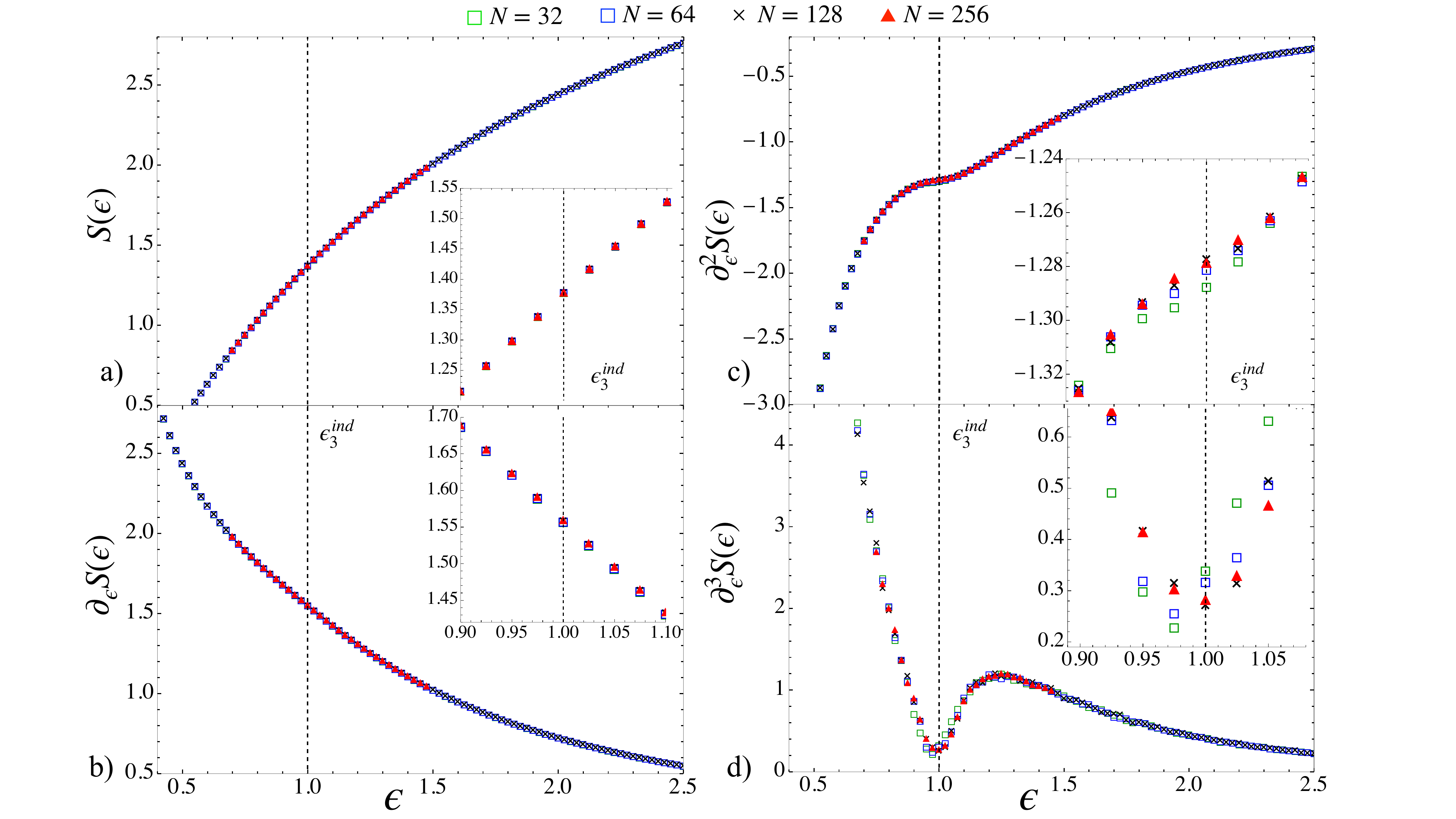}
    \caption{\textbf{Microcanonical entropy and its energy derivatives for the 2D-sG model in the thermodynamic limit}. The black dashed line in all the plots represents the transition energy value $\epsilon_{3}^{ind}=1.0$. Plot \textbf{a)} Entropy function versus energy density $\epsilon$. The inset shows the behavior of $S(\epsilon)$ around the ``critical'' energy. Plot \textbf{b)} First-order derivative of entropy versus energy density $\epsilon$. The inset shows the behavior of $\partial_{\epsilon}S(\epsilon)$ in the ``critical'' region around $\epsilon_{3}^{ind}$. Plot \textbf{c)} Second-order derivative of entropy versus energy density $\epsilon$. The inset shows the behavior of $\partial^2_{\epsilon}S(\epsilon)$ around the ``critical'' region where such a function admits an inflection point in correspondence to the black dashed line. Both $\partial_{\epsilon}S(\epsilon)$ and $\partial^2_{\epsilon}S(\epsilon)$ are computed numerically using the PHT method (see Appendix~\ref{app:PHT_method}). Plot \textbf{d} Third-order derivative of entropy computed with the finite difference (central) method from $\partial^2_{\epsilon}S(\epsilon)$. The inset shows the positive-valued minimum of $\partial^3_{\epsilon}S(\epsilon)$ that represents the fingerprint of a third-order PT according to MIPA. Note that the minimum changes depending on the system size, i.e., $\epsilon^{ind,N=64}_{3}\approx 0.98$ (see red triangle) and stabilizes around $\epsilon^{ind}_{3}= 1.0$ for $N\geq128$ (see the black crosses and the red triangles).}
      \label{fig:derivatives_entropy_thermo_limit}  
\end{figure*}
\subsubsection*{Advantages of adopting the microcanonical ensemble} Switching to the microcanonical ensemble to investigate PTs is not merely propaedeutic to the implementation of MIPA. It is necessary to solve conceptual problems like the inequivalence of ensembles that usually hinder RG methods. 
In fact, there exists a large class of both quantum and classical systems that interact through long-range potentials, for which the RG theory is not readily applicable. In this case, it is well known that long-range interactions can give rise to the non-equivalence of the canonical and microcanonical ensemble~\cite{barre2001inequivalence, lynden1968gravo, dauxois2000violation, kiessling1997micro}. 

Rigorous results show that the equivalence of the ensembles is lost when a first-order (discontinuous) PT occurs \cite{touchette2004introduction, touchette2003equivalence,ellis2002nonequivalent, touchette2005nonequivalent, ellis2004thermodynamic}. In this case, while microcanonical observables can always be defined, canonical ones may suffer mathematical pathologies \cite{touchette2005nonequivalent}.\\
Therefore, the advantage of adopting the microcanonical ensemble lies in providing a theoretical framework that is more fundamental than the canonical one \cite{callaway1983lattice}.

\subsubsection*{Microcanonical classification of PTs}

For this purpose, Bachmann et al. have developed the microcanonical inflection point analysis (MIPA) \cite{bachmann2014thermodynamics,qi2018classification,sitarachu2022evidence} as a result of a long and deep study initiated by Gross et al. \cite{gross2001microcanonical,gross2005microcanonical,gross2002geometric,gross2001ensemble,matty2017comparison,chomaz1999energy,dunkel2006phase,hilbert2014thermodynamic}. 

MIPA consists in the analysis of the energy-behaviors of fundamental microcanonical observables such as entropy and its energy derivatives, say $\partial^n_{E}S(E)$, for some integer $n$. Briefly, MIPA recognizes two classes: independent and dependent PTs. Independent PTs represent the major change in the system's phase, whereas a dependent PT can only occur (but not necessarily) if an independent one already manifested. They are considered precursors of a major transition. Focusing on the first class, an independent PT of even (odd) order $2n$ ($2n-1$) occurs if $\partial^{2n-1}_{E}S$ ($\partial^{2n-2}_{E}S$) admits an inflection point at $E_c$ that gives rise to a negative-valued maximum (positive-valued minimum) in $\partial^{2n}_{E}S$ (in $\partial^{2n-1}_{E}S$). MIPA represents a classification scheme for PTs in the microcanonical ensemble. For instance, an independent second-order PT is detected by the presence of an inflection point at $E_c$ in the energy profile of $\partial_{E}S$, while $\partial^{2}_{E}S$ admits a negative-valued maximum in $E_c$. \\
An independent third-order PT is detected by an inflection point at $E_c$ in the energy profile of $\partial^2_{E}S$, while $\partial^3_{E}S$ admits a positive-valued minimum in $E_c$.

In order to compute the microcanonical observables for applying MIPA, it is necessary to build the microcanonical ensemble. Therefore, we introduce the Hamiltonian function for the 2D-sG theory. This can be done by associating a momentum $\pi_{\bm{n}}$ to each field's degrees of freedom $\phi_{\bm{n}}$; this procedure naturally defines the kinetic energy
$K[\pi]:=\sum_{\bm{n}}\pi^{2}_{\bm{n}}/2$ that is used to define the Hamiltonian function as $H[\pi,\phi]:=K[\pi]+V_{sG}[\phi]$. In so doing, the microcanonical entropy function, $S(E)$, is then introduced based on the microcanonical partition function $\Omega(E)$ and they read ($k_B=1$)
\begin{equation}
    S(E)=\log \Omega(E),\quad\Omega(E):=\int \delta(H[\pi,\phi]-E)\;D\pi\,D\phi\,\,.
\end{equation}
Entropy is the fundamental thermodynamic potential that generates all microcanonical (thermodynamic) observables as does free energy in the canonical ensemble. More precisely, any thermodynamic observable can be obtained by differentiating the entropy function with respect to the energy variable (which is considered fixed in the microcanonical ensemble). 

To have access to these observables, we carry out numerical simulations in the microcanonical ensemble through the method proposed in Refs.~\cite{ray1996microcanonical,ray1991microcanonical}. The energy derivatives of entropy are numerically computed using the Pearson-Halicioglu-Tiller (PHT) method (see Appendix~\ref{app:micro_montecarlo}).

\section{Results}

We perform an intensive numerical investigation by computing the aforementioned thermodynamic observables and applying the microcanonical analysis. We follow two directions. \\
In the first, we fix the values of the parameters, choosing $g=3$ and $K=3$, and study the microcanonical observables and the phenomenology of this system for different lattice sizes $(N= 32,\, 64,\, 128,\, 256)$, thus mimicking the thermodynamic limit. \\
In the second one, we have explored the thermodynamic properties of the sG model at fixed lattice size $N=32$ but different values of the system's parameters $(g,K)$. This analysis is reported in Appendix~\ref{app:mipa_fixed_size}.

\subsection{Microcanonical analysis in the limit of increasing system size}

In Fig.~\ref{fig:derivatives_entropy_thermo_limit}\textbf{a)}, we report the first-order derivative of entropy $\partial_{\epsilon}S(\epsilon)=T^{-1}(\epsilon)$ that corresponds to the inverse temperature. At any size, the inverse temperature is continuous and does not admit any inflection point. According to MIPA, neither a first- nor a second-order PT is manifesting. This behavior does not exclude the emergence of higher-order PTs that are usually not deeply studied in the literature. 

To see that, we compute the second-order derivative ($\partial^{2}_{\epsilon}S(\epsilon)$) and third-order ($\partial^{3}_{\epsilon}S(\epsilon)$) derivative of the entropy, reported in Fig.~\ref{fig:derivatives_entropy_thermo_limit}\textbf{b)} and \ref{fig:derivatives_entropy_thermo_limit}\textbf{c)}, respectively. The function $\partial^{2}_{\epsilon}S(\epsilon)$ admits an inflection point around $\epsilon\approx 1.0$ (see the black dashed line) that gives rise to a positive-valued minimum in $\partial^{3}_{\epsilon}S(\epsilon)$. According to MIPA, this behavior signals the presence of an independent third-order PT that we identify with \emph{roughness PT}. 

Several observations are now in order. First, it should be stressed that the value of the minimum varies with the size of the system but reaches a stable value for $N\geq128$ (see the inset of Fig.~\ref{fig:derivatives_entropy_thermo_limit}\textbf{c)}. In fact, at $N=64$, the minimum of $\partial^{2}_{\epsilon}S(\epsilon)$ is reached at $\epsilon^{ind,N=64}_{3}\approx0.98$, whereas for $N\geq 128$ the minimum converges to $\epsilon\approx 1.0$. From now on, we denote the third-order transition energy by $\epsilon^{ind}_{3}=1$. Second, according to MIPA, we note that the positive-valued maximum in $\partial^{3}_{\epsilon}S(\epsilon)$ around $\epsilon\approx 1.25$ does not correspond to any PT. Moreover, by inspection of Fig.~\ref{fig:derivatives_entropy_thermo_limit}\textbf{c)}, $\partial^{3}_{\epsilon}S(\epsilon)$ changes visibly concavity from down to up around $\epsilon\approx 1.5$, but such a point gives rise to a negative-valued minimum in $\partial^{4}_{\epsilon}S(\epsilon)$, so this point does not correspond to any PT.

Finally, we discuss a remarkable feature of higher-order PTs, namely, the absence of non-analyticities. Observing Figs.~\ref{fig:derivatives_entropy_thermo_limit}\textbf{a)}-\textbf{c)}, all of the derivatives of entropy, in particular, around the region where the third-order PT occurs, no evidence of catastrophic behaviors or evolution toward divergences is observed. In fact, the function $\partial^{3}_{\epsilon}S(\epsilon)$ remains rounded at $\epsilon_{3}^{ind}\approx 1.0$, and the value of its minimum is still stable after doubling the size of the system. \\
Only recently, this behavior began to be noticed in some systems at increasing sizes, such as in the 4D-U(1) lattice gauge theory \cite{di2024detecting}, and in the Ising model \cite{sitarachu2020exact}, but also at infinite size \cite{sitarachu2022evidence}.

\subsection{Specific heat anomaly}\label{ssec:specific_heat}

\begin{figure}[tbh!]
		\centering
		\includegraphics[height=11cm, width=8cm,keepaspectratio]{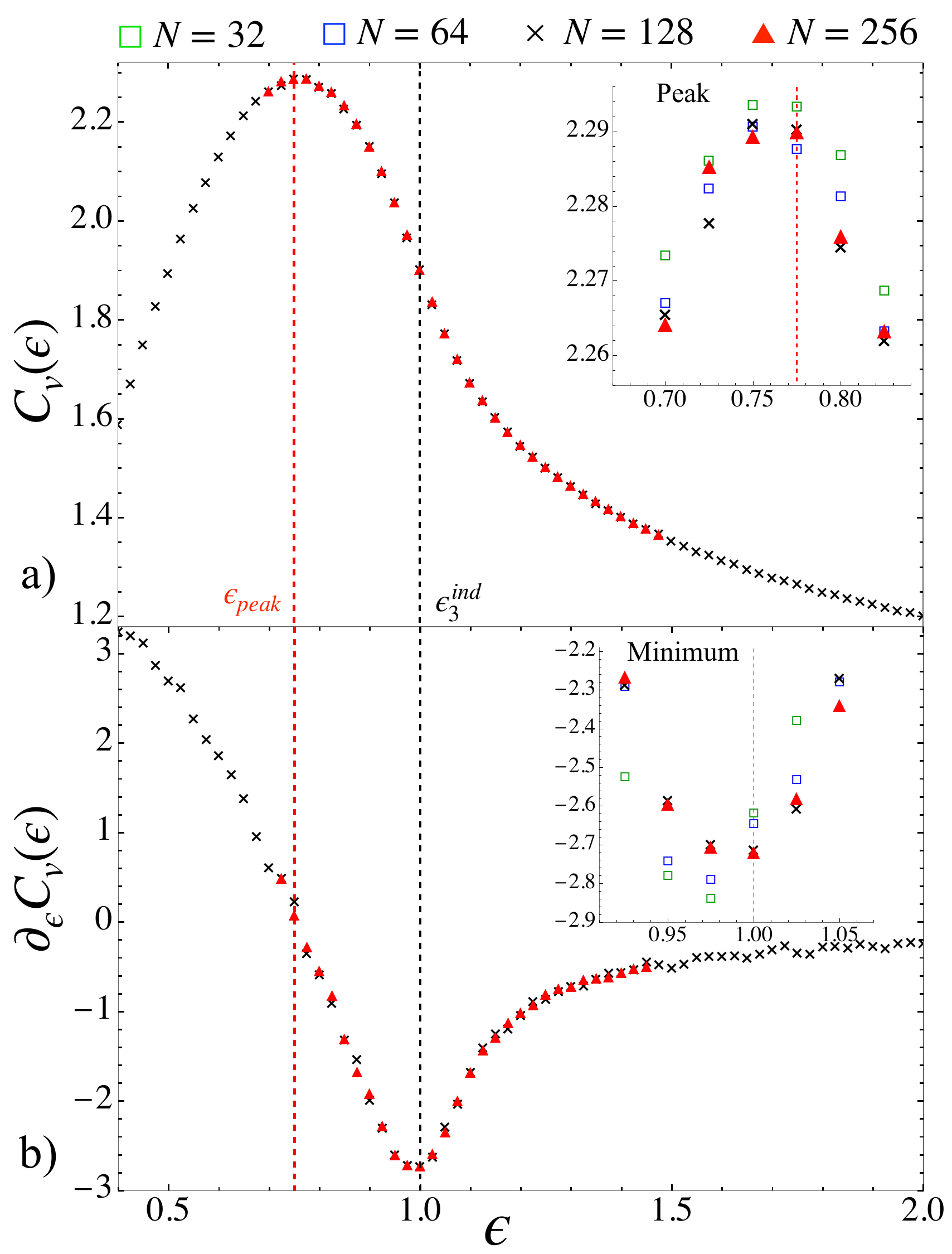}

		\caption{\textbf{Specific heat and its energy derivative in the microcanonical ensemble versus the specific energy}. Plot \textbf{a)} displays the specific heat $C_v(\epsilon)$ for $N=128$ (black crosses) and $N=256$ (red crosses). The vertical black dashed line represents the transition energy, $\epsilon^{ind}_{3}=1.0$ while the vertical red dashed line locates the maximum in the specific heat $\epsilon_{peak}=0.775$. Plot \textbf{b)} displays the energy derivative of the specific heat $\partial_{\epsilon}C_v(\epsilon)$, computed through the finite difference (central) method. In the inset of plot \textbf{a)} a zoom on the peak of the specific heat is reported for all the lattice sizes studied in this work as indicated by the legend above. Note that the values of the peak remain confined between in the range $[2.285,2.290]$ for increasing system sizes. $C_v(\epsilon)$ admits an inflection point in correspondence with $\epsilon^{ind}_{3}=1.0$. In plot \textbf{b)}, we report $\partial_{\epsilon}C_v(\epsilon)$ for the same sizes as in plot \textbf{a)}. Similarly, in the inset we report a zoom on the minimum of $\partial_{\epsilon}C_v(\epsilon)$ and we show a comparison between all the lattice sizes as shown in the legend. This proves that the specific heat contains information on the location of the transition energy but the extrapolation is dictated by the MIPA paradigm.} 
	\label{fig:specific_heat}
\end{figure}

In this section, we study the specific heat for the 2D-sG model showing the emergence of the SHA. \\
We thus show how to detect a third-order PT by analyzing the energy profile of the specific heat. Then, we show how to reconcile the standard approach for detecting PT based on the specific heat with MIPA method. This analysis shows that the roughness PT can be detected from the specific heat and it thus corresponds to the minimum of $\partial_{E}C_v(E)$. \\
The specific heat in the microcanonical ensemble is defined by:
\begin{equation}\label{eqn:specific_heat}
    C_v(E)=-\frac{(\partial_{E}S(E))^2}{\partial^2_{E}S(E)}\,.
\end{equation}
The energy behavior of $C_v$ and $\partial_{E}C_v$, obtained from the numerical data, are plotted in Figs.~\ref{fig:specific_heat}\textbf{a)} and ~\ref{fig:specific_heat}\textbf{b)}, respectively. Observing the inset of Fig.~\ref{fig:specific_heat}\textbf{a)}, we can notice that $C_v$ admits a rounded peak and that the maximum does not appreciably vary its numerical value at increasing lattice size. Furthermore, note that the maximum of the specific heat is located around $\epsilon_{peak}= 0.775$, see the vertical red dashed line in Fig.~\ref{fig:specific_heat}\textbf{a)}. It is notably distant from the transition energy predicted by MIPA. 

Further inspecting Fig.~\ref{fig:specific_heat}\textbf{a)}, we notice that $C_v$ admits an inflection point at $\epsilon^{ind}_{3}\approx 1.0$ which yields a minimum in $\partial_{E}C_v(E)$, see Fig.~\ref{fig:specific_heat}\textbf{b)}. Thus, in the presence of the specific heat anomaly, one can detect the transition by searching for the inflection point of $C_v(E)$ and then the minimum of $\partial_{E}C_v(E)$. 

From the conceptual viewpoint, it should be stressed that such a procedure is consistent with the fact that the transition is of the third-order. In fact, the specific heat is an observable of \textit{second-order}, for it is proportional to second-order fluctuations of energy $C_v\approx \langle E^2\rangle-\langle E\rangle^2$, then $\partial_{E}C_v(E)$ encodes third-order fluctuations. Moreover, the fact of having a non-diverging peak is fully justified by the behavior of $\partial_{E}^2S(E)$ as we deduce from the following reason. 

If the transition was of the second-order, then $\partial_{E}^2S(E)$ would admit a (negative-valued) maximum that moves to zero, in the thermodynamic limit (see, for instance, Fig.~1 in Ref.~\cite{sitarachu2022evidence}). In so doing, according to Eq.~\eqref{eqn:specific_heat} $C_v(E)$ would admit a peak that diverges in the same limit.

In our case, instead, $\partial_{E}^2S(E)$ admits an inflection point that does not \textit{move} in the infinite size limit (see again Figs.~\ref{fig:derivatives_entropy_thermo_limit}\textbf{a)} and \ref{fig:derivatives_entropy_thermo_limit}\textbf{b)}), thus leaving the maximum of the specific heat quantitatively constant. \\
We thus conclude that the presence of a constant, non-diverging peak in the specific heat is the fingerprint of the emergence of a third-order PT. This is the mechanism that should manifest also in other systems such as the XY-model.

\subsection{Characterization and phenomenological aspects}

The thermodynamic analysis provided by MIPA has revealed that the roughness PT is of third-order and that it is just such a third-order feature to give rise to the characteristic energy profile of the SHA. In this subsection, we study the phenomenological aspect of the roughness PT and its characterization in terms of physical observables. 

The scope is twofold. Our first goal is to find a pattern associated with the roughness transition that is unbiased with respect to MIPA. We thus study the histograms associated with the field configurations and two types of ``order parameters'' such as the \emph{roughness} and \emph{field hierarchy} parameters. They allow for a better understanding of the mechanism that triggers the transition itself. Our second scope is to provide physical evidences confirming that the peak of the specific heat does not correspond to any independent transitional phenomenon. 

\subsubsection{Histograms of field configurations}

\begin{figure*}[tbh!]
\centering
    \includegraphics[height=15cm, width=17.5cm,keepaspectratio]{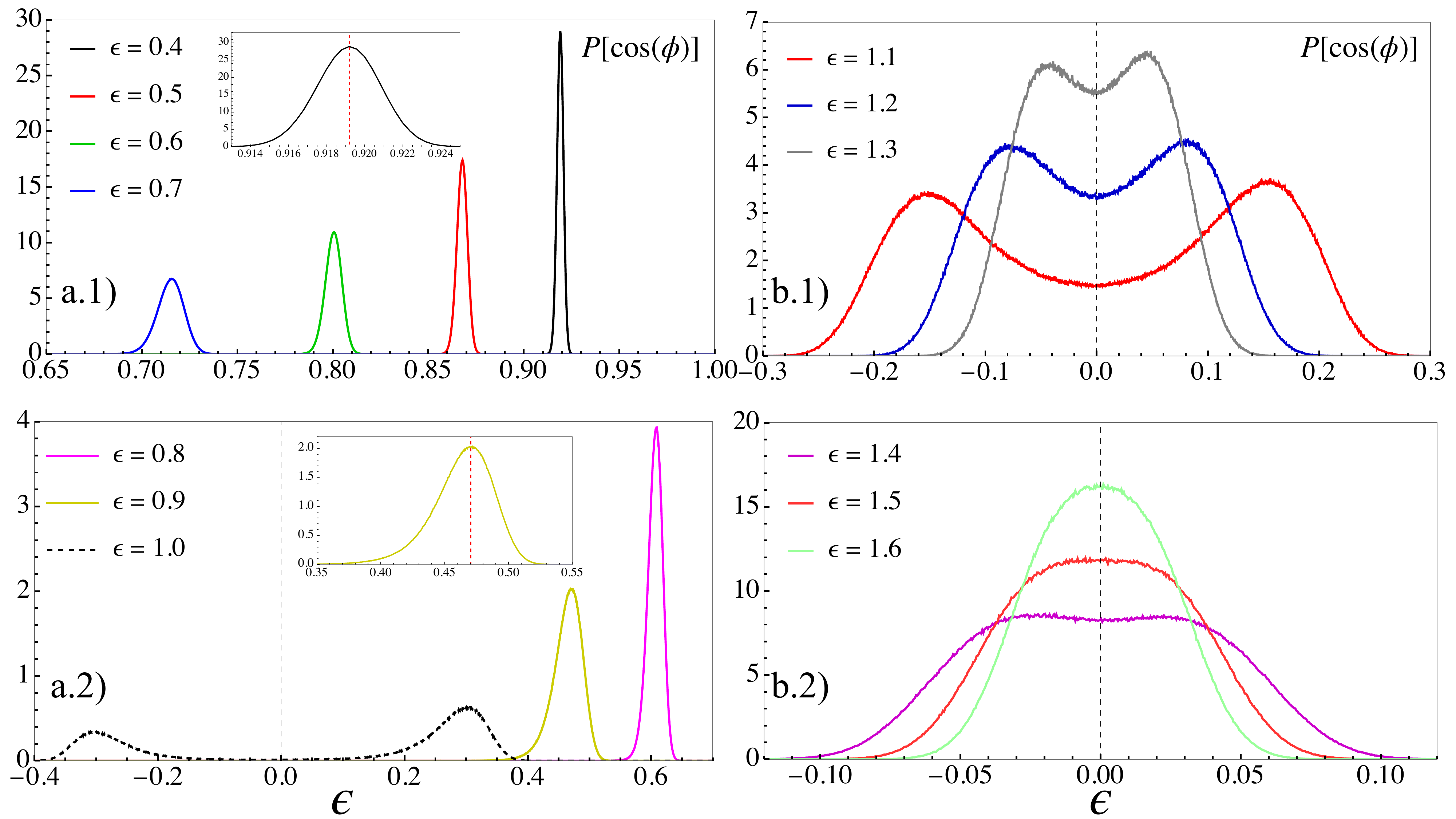}
    \caption{\textbf{Histograms of the field values $\cos(\phi)$ for $N=128$}. Due to the periodicity of the cosine function the range of possible field configurations coincides with the interval $[-1,1]$.  In plot \textbf{a.1)}, $P[\cos(\phi)]$ are displayed at low energy values, i.e., $[0.4,0.7]$. In this energy regime, the histograms follow a Gaussian shape as shown in the inset where $P[\cos(\phi)]$ at $\epsilon=0.4$ is plotted and it is evidently symmetric with respect to the vertical dashed red line. In plot \textbf{a.2)}, the histograms for energy values $\epsilon=0.8$ and $0.9$ acquire a left tail and, as shown in the inset, the histogram become asymmetric. Then, at $\epsilon=1.0$ (see the dashed black histograms), a double peak appears in $P[\cos(\phi)]$. Then, in plot \textbf{b.1)}, we observe that these peaks begin merging for increasing temperature and finally, in plot \textbf{b.2)}, see the green histogram, the Gaussian shape is reached again.}
      \label{fig:histogram_cosphi}  
\end{figure*}

\begin{figure*}[tbh!]
    \includegraphics[height=10cm, width=18cm,keepaspectratio]{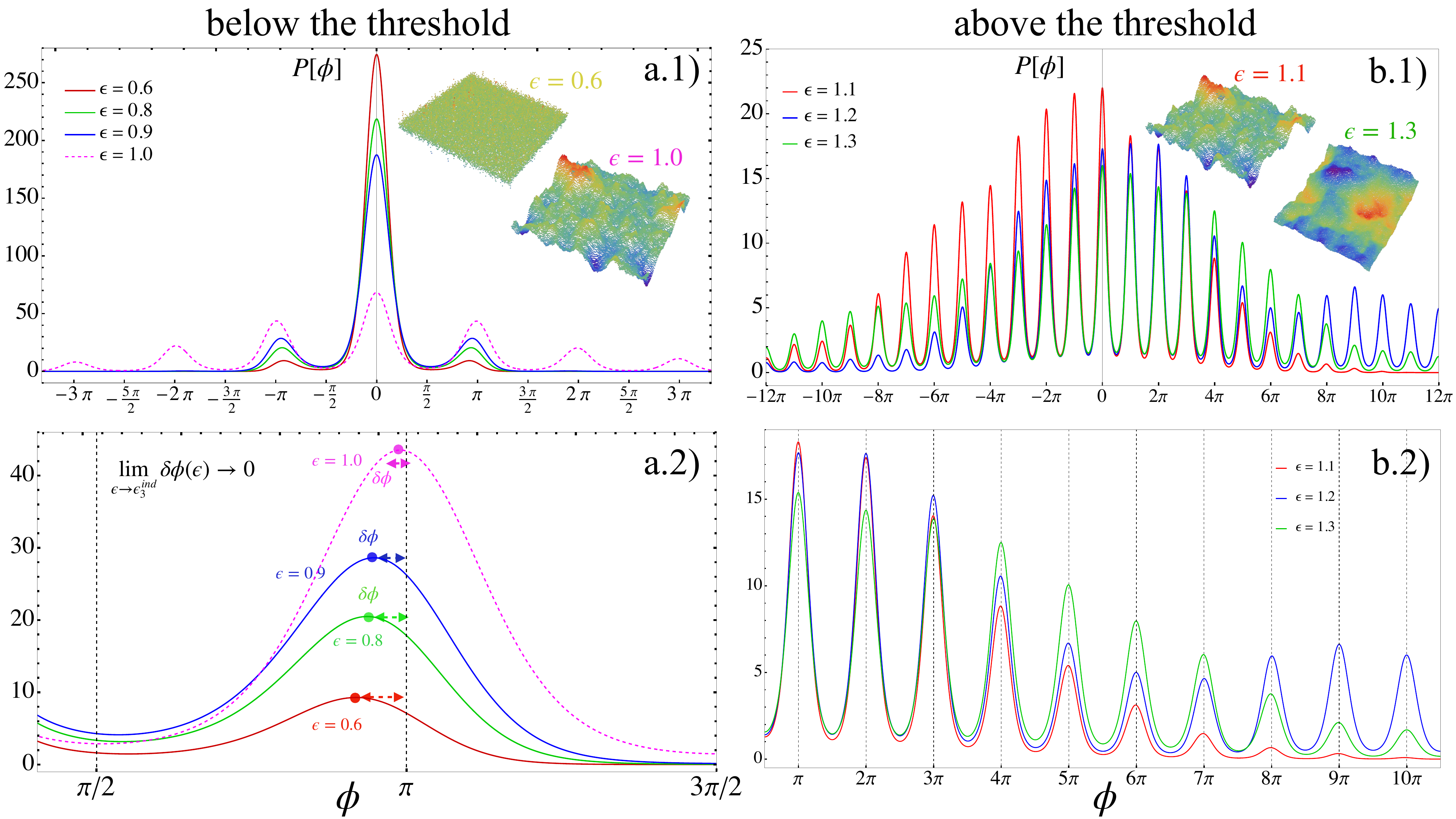}
    \caption{\textbf{Histograms of the field variable $\phi$ below and above the transition energy, $\epsilon^{ind}_{3}=1.0$, for lattice size $N=128$}. All the plots represent the (non-normalized) histograms $P[\phi]$ regarded as the frequency of occurrence of the field configurations. Hence, $P[\phi]$ counts the number of occurrences of the value occupied by the field $\phi_{\bm{n}}$ for any lattice site $\bm{n}\in\mathbb{L}$. Plot \textbf{a.1)} displays the histograms for specific energy values below the transition energy $\epsilon_{3}^{ind}$ while plot \textbf{a.2)} is a zoom of plot \textbf{a.1)} on the peak located at $\phi=\pi$. In plot \textbf{a.1)}, we report some snapshots of a (thermalized) lattice configuration obtained through the statistical average of each lattice field variable $\langle \phi_{\bm{n}}\rangle(E)$. Notice that all the $P[\phi]$ below $\epsilon_{3}^{ind}$ admit a highest peak at $\phi=0$ and smaller peaks at $\phi=n\pi$ for increasing values of the integer $n$.\\
    Plot \textbf{b.1)} displays the histograms of field configuration at specific energy values above the transition energy $\epsilon_{3}^{ind}$. Plot \textbf{b.2)} shows a zoom of plot \textbf{b.1)} on several peaks. Note that these peaks now are smeared throughout the range of possible configurations and they exactly fall onto many multiples of $\pi$. In plot \textbf{b.1)}, we report some snapshots of a (thermalized) lattice configuration obtained through the statistical average of each lattice field variable $\langle \phi_{n}\rangle(E)$.}
      \label{fig:histogram_phi_above_below_thresh}  
\end{figure*}

We study the histograms $P[\cos(\phi)]$ and $P[\phi]$ associated, respectively, with the cosine of field configurations and with the field configurations at different energy values. \\
\textbf{Below the energy threshold}, $\epsilon^{ind}_{3}$. Histograms $P[\cos(\phi)]$. These histograms are shown in Figs.~\ref{fig:histogram_cosphi}\textbf{a.1)} and \ref{fig:histogram_cosphi}\textbf{a.2)}. At very low energy, histograms $P[\cos(\phi)]$ have a Gaussian profile centered on some positive value between $[0,1]$. At very low energies, see the black curve in Fig.~\ref{fig:histogram_cosphi}\textbf{a.1)}, $P[\cos(\phi)]$ have a small width and a large maximum value. As soon as the energy increases, the maximum decreases, and the width of the Gaussian profile becomes larger. \\
Not only, the histograms start deviating from Gaussian behavior and a left tail arises (asymmetry); see the yellow histogram in the inset of Fig.~\ref{fig:histogram_cosphi}\textbf{a.2)}. At the transition energy, the histograms develop a double peak symmetrically centered on a positive and negative value of the domain $[-1,1]$; see the dashed black histogram Fig.~\ref{fig:histogram_cosphi}\textbf{a.2)}. Histograms
$P[\phi]$. They consist of periodic peaks located on multiples of $\pi$, i.e. $\phi_{\pm n}=\pm n\pi$ with integer $n$. The highest peak is present at $\phi=0$, while the peaks on $\phi_{\pm n}$ are successively smaller for increasingly larger values of $n$, that is, $P[\phi_0]>P[\phi_1]>P[\phi_2]>\ldots$. At very low energies, $P[\phi]$ admit very few peaks, $\phi_0,\,\phi_{\pm 1}$, and $P[\phi_0]>>P[\phi_{\pm1}]$, that is, the central peak is very large with respect to the other; see the red curve for $\epsilon=0.6$ in Fig.~\ref{fig:histogram_phi_above_below_thresh}\textbf{a.1)}. 

For increasing values of $\epsilon$, new peaks emerge in $|n|>1$ and the central peak in $\phi_0$, which was the highest one, now is comparable with the others $P[\phi_0]\approx P[\phi_{\pm 1}]$; see the dashed magenta curve for $\epsilon=1.0$ in Fig.~\ref{fig:histogram_phi_above_below_thresh}\textbf{a.1)}. Moreover, we notice that $P[\phi]$ have minima around $\phi=\pm\pi/2,\,\pm 3\pi/2,\ldots$ and, although $P[\pm\pi/2]$ is very small, it does not vanish. For instance, for $\epsilon=0.6$, we have $P[0]\approx 270$, $P[\pi]\approx 10$, and $P[\pi/2]\approx 2$ then, $P[\pi/2]/P[0]\approx 0.0074$.

From a phenomenological perspective, combining all the information together, the behaviors of $P[\cos(\phi)]$ and $P[\phi]$ indicate that the field configurations are mostly distributed around a specific value around $\phi\approx\phi_0$ that, in principle, would produce a peak in $P[\cos(\phi)]$ around $\cos(\phi)~\approx 1$. However, other configurations around $\phi_{\pm 1}$ are allowed; therefore, the Gaussian histogram peak $P[\cos(\phi)]$ is shifted at some value (smaller than 1) of a quantity proportional to the height of $P[\phi_{\pm 1}]$. \\
For energies close to the transition one, we have $P[\phi_0]\approx P[\phi_{\pm 1}]$, which implies that $P[\cos(\phi_0)]\approx P[\cos(\phi_1)]$ thus produces a double peak symmetrically centered around $\cos(\phi)=0$.\\ Clearly, the peaks are not centered exactly on $\pm1$, since the contribution of intermediate states between $\phi_{n}$ and $\phi_{n+1}$ is not negligible and contributes to the shift of the peaks from this extreme configuration.\\
\textbf{Above the energy threshold}, $\epsilon^{ind}_{3}$. Histogram $P[\cos(\phi)]$. They are reported in Figs.~\ref{fig:histogram_cosphi}\textbf{b.1)} and~\ref{fig:histogram_cosphi}\textbf{b.2)}. The double peaks start increasing their numerical value and their tails merge; see the red curve for $\epsilon=1.1$ in Fig.~\ref{fig:histogram_cosphi}\textbf{b.1)}. For increasing values of energy, the two peaks merge together (magenta curve in Fig.~\ref{fig:histogram_cosphi}\textbf{b.2)}) thus reaching a Gaussian profile at relatively high energies (green curve in Fig.~\ref{fig:histogram_cosphi}\textbf{b.2)}).\\
Histograms $P[\phi]$. These histograms are shown in Fig.~~\ref{fig:histogram_phi_above_below_thresh}\textbf{b.1)}. They develop a large number of peaks with comparable probability. For example, see the histograms for $\epsilon=1.1$ in Fig.~\ref{fig:histogram_phi_above_below_thresh}\textbf{b.1)} $P[0]\approx P[-\pi]$, and in Fig.~\ref{fig:histogram_phi_above_below_thresh}\textbf{b.2)}, $P[\pi]\approx P[2\pi]$. Moreover, the values of $P[\phi]$ do not vanish around $\phi=\pm \pi/2,\,\pm 3\pi/2,\ldots$ but now the ration $P[0]/P[\pi/2]$ is larger than below the transition temperature. For example, for $\epsilon=1.1$, we have $P[\pi/2]\approx 1$, and $P[\pi/2]/P[2\pi]\approx 0.05$.

Phenomenologically, this is the effect that causes the merging of the tails in the double peak of the histograms $P[\cos(\phi)]$. Another interesting mechanism emerges at the transition energy. This is shown in Fig.~\ref{fig:histogram_phi_above_below_thresh}\textbf{a.2)}. We see that a low energies the second-highest peak does not exactly fall onto $\phi_{\pm 1}=\pm \pi$, instead, we observe a shift $\delta\phi=\phi_{peak}-\pi$. This shift decreases with energy, converging to $\delta\phi(\epsilon)\to 0$ as $\epsilon\to\epsilon^{ind}_{3}$.

\subsubsection{Roughness and field hierarchy parameters}

\begin{figure}[tbh!]
    \includegraphics[height=12cm, width=8cm,keepaspectratio]{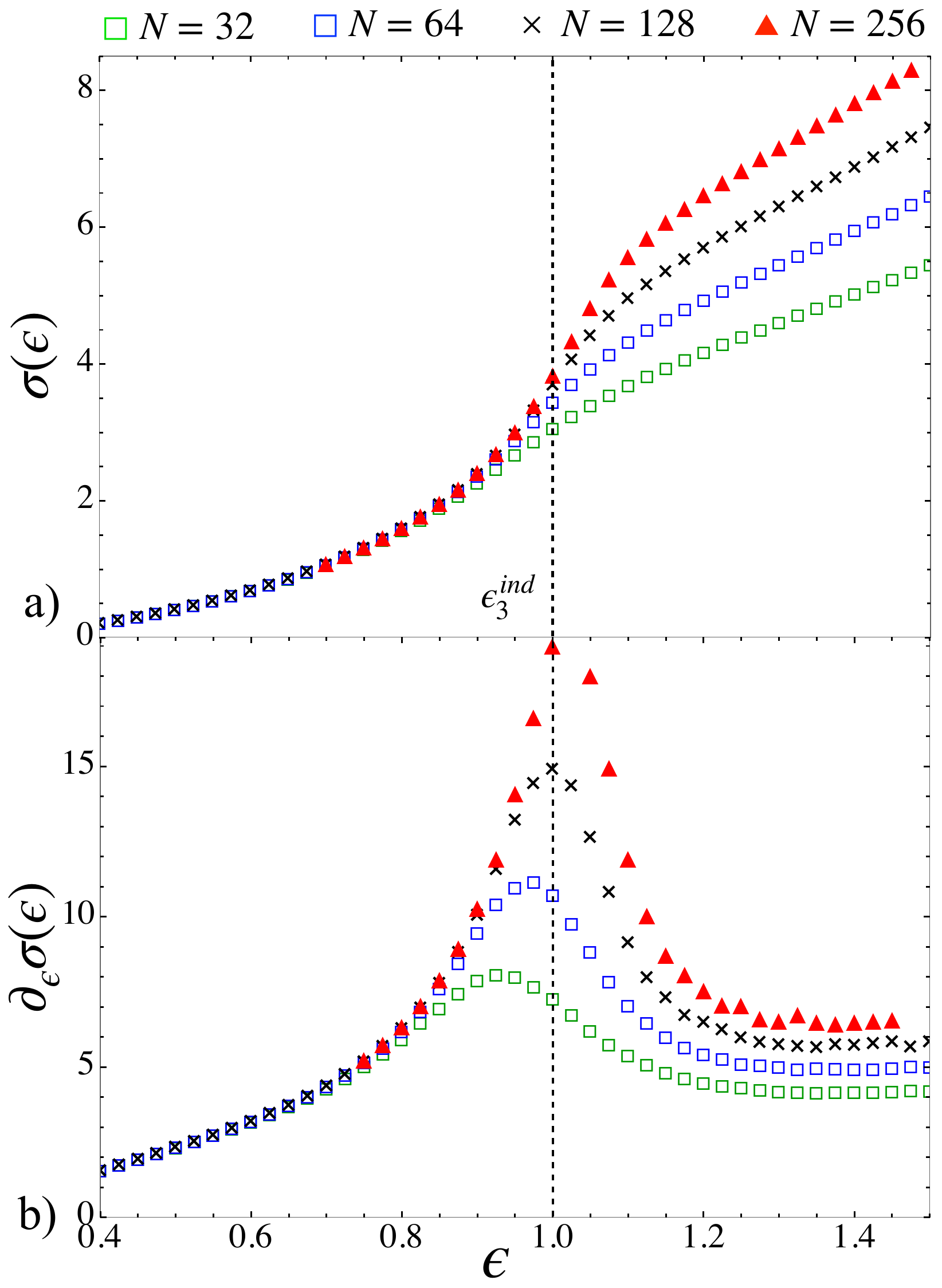}
    \caption{\textbf{Roughness parameter and its energy derivative versus specific energy at different system sizes}. The vertical black dashed line localizes the transition energy $\epsilon^{ind}_{3}=1.0$. Plots \textbf{a)} shows a comparison of $\sigma(\epsilon)$ for $N=32,\,64,\,128,\,256$ as reported by the legend above. An inflection point is visible at $\epsilon^{ind}_{3}$. In plot \textbf{b)}, the energy derivative of the roughness parameter, $\partial_{\epsilon}\sigma(\epsilon)$, is reported for $N=32,\,64,\,128,\,256$. Note that a pronounced maximum emerges in correspondence with $\epsilon^{ind}_{3}$. This proves a robust pattern between MIPA and the phenomenology encoded in the roughness parameter.}
      \label{fig:roughness_parameters}  
\end{figure}
\begin{figure}[tbh!]
    \includegraphics[height=12cm, width=8.3cm,keepaspectratio]{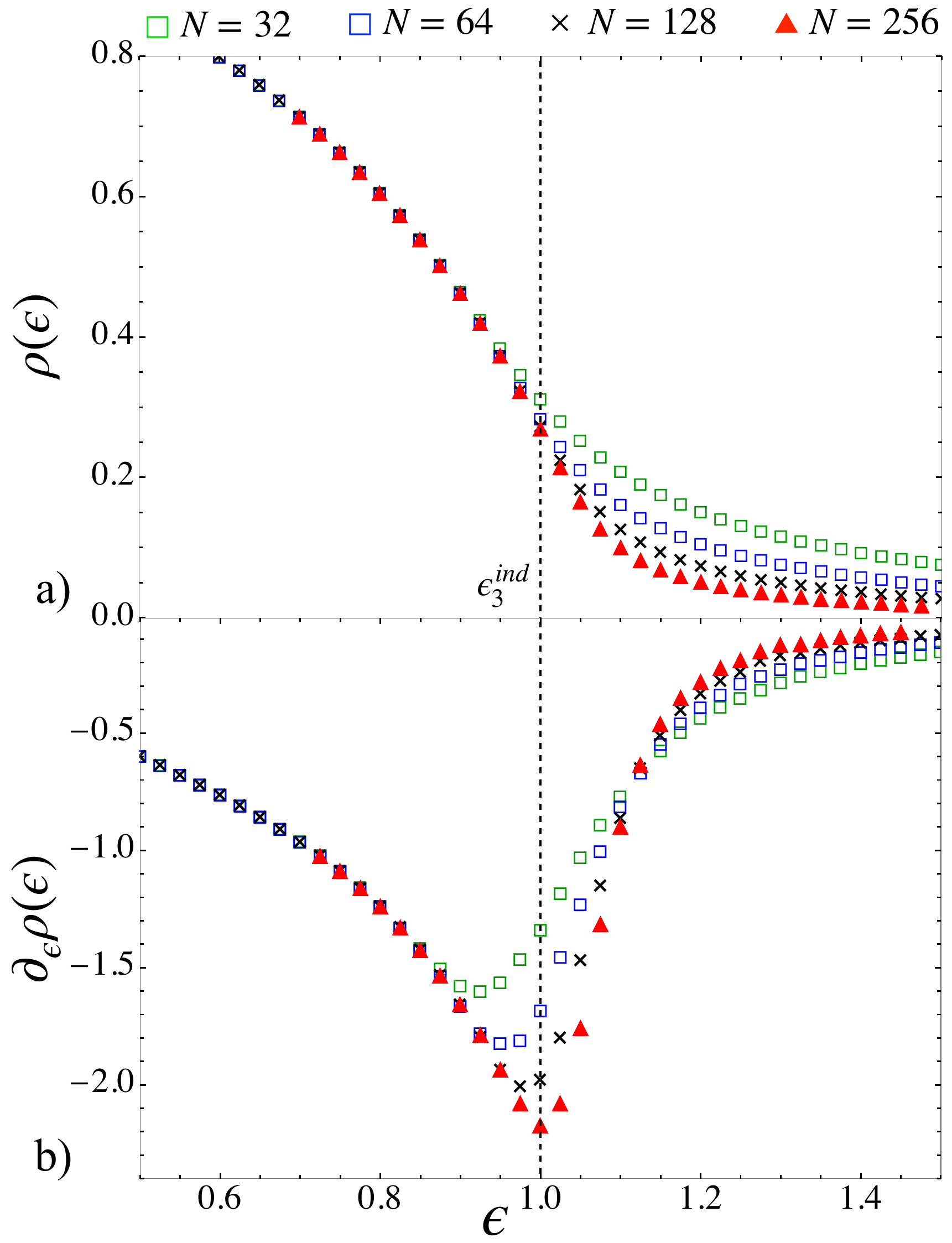}
    \caption{\textbf{Field hierarchy parameter and its energy derivative versus specific energy at different system sizes}. The vertical black dashed line localizes the transition energy $\epsilon^{ind}_{3}=1.0$. Plot \textbf{a)} shows a comparison of $\rho(\epsilon)$ for all the system sizes reported in the legend above. An inflection point is visible at $\epsilon^{ind}_{3}$. In plot \textbf{b)}, the energy derivative of the field hierarchy parameter, $\partial_{\epsilon}\rho(\epsilon)$, is reported for $N=32,\,64,\,128,\,256$ as explained by the legend above. Note that a pronounced minimum emerges in correspondence with $\epsilon^{ind}_{3}$. This proves another robust pattern between MIPA and the phenomenology encoded in the field hierarchy parameter.}
      \label{fig:hierarchy_parameter}  
\end{figure}

The properties of the histograms $P[\phi]$ and $P[\cos(\phi)]$ can be condensed into a few scalar observables. In so doing, we can better capture the signal of the third-order PT detected by MIPA. \\
We thus introduce the quantity $\Phi:=\frac{1}{N}\sum_{i\in\mathbb{L}}\phi_{i},$
that allows us to define the \textit{roughness parameter} \cite{ares2004equilibrium,sanchez2000roughening}
\begin{equation}
    \sigma(E) :=\frac{1}{N}\sum_{i\in\mathbb{L}}\langle (\phi_i - \Phi)^2\rangle_{E}\;.
\end{equation}
Then, we introduce the vector $\mathbb{C}:=(\mathbb{C}_x,\mathbb{C}_y)$ such that $   \mathbb{C}_x:=\frac{1}{N}\sum_{i\in\mathbb{L}} \cos(\phi_i)$ and $\mathbb{C}_y:=\frac{1}{N}\sum_{i\in\mathbb{L}} \sin(\phi_i)$.
In so doing, we define the \textit{field hierarchy parameter}:
\begin{equation}
\begin{split}
    \rho(E):=\langle\|\mathbb{C}\|\rangle_{E}=\Big\langle\sqrt{\mathbb{C}_{x}^{2}+\mathbb{C}_{y}^2}\Big\rangle_{E}\,,
\end{split}
\end{equation}
where the notation $\langle\cdot\rangle_{E}$ stands for the microcanonical statistical average. \\
We remark that, due to the third-order feature of the roughness PT, we expect that observables of lower-than-second order are weakly sensible to such a transition. For that reason, the strategy is to introduce second-order observables and then to differentiate them so as to enhance relevant (third-order) behaviors around the transition energy. Hence, in Fig.~\ref{fig:roughness_parameters} \textbf{a)}, we see that $\sigma$ has an inflection point exactly at $\epsilon^{ind}_{3}= 1.0$ and $\partial_{\epsilon}\sigma$ admits a maximum. This behavior confirms that there is no pattern with the peak in $C_v$. 

Therefore, the pattern is associated with the minimum of $\partial^{3}_{\epsilon}S$ as correctly predicted by MIPA. Then, in Fig.~\ref{fig:hierarchy_parameter}\textbf{a)} and \textbf{b)}, we also report the field hierarchy parameter, $\rho(\epsilon)$ and its energy derivative, $\partial_{\epsilon}\rho(\epsilon)$, respectively. \\
Yet, we observe an analog behavior as for the roughness parameter. Indeed, $\rho(\epsilon)$ admits an inflection point in $\epsilon^{ind}_{3}$ that produces a minimum (negative peak) in its $\partial_{\epsilon}\rho(\epsilon)$. This behavior confirms that the MIPA provides the correct thermodynamic information, namely, that the 2D-sG model undergoes a third-order PT corresponding to the roughness transition, and unbiased observables with respect to MIPA are consistent with MIPA itself.

\section{Conclusions}

The findings of this work have several implications that deserve careful discussion. 

Using MIPA, we have detected a third-order PT that we identified with the roughness PT, as it occurs at the energy where the roughness parameter $\rho(\epsilon)$ exhibits a peak and $\partial_{\epsilon}\rho(\epsilon)$ shows a minimum; see Figs.~\ref{fig:roughness_parameters}\textbf{a)} and ~\ref{fig:roughness_parameters}\textbf{b)}. Moreover, the transition energy $\epsilon^{ind}_{3}=1$ associated with the third-order PT is well separated from the energy $\epsilon_{peak}=0.775$ where the peak in the specific heat appears. MIPA detects no inflection point in the entropy derivative around $\epsilon_{peak}$, indicating that this behavior is not associated with any PT. 

Instead, the SHA can be understood as a consequence of the emergence of the roughness PT at $\epsilon^{ind}_{3}$. As discussed in section~\ref{ssec:specific_heat}, in the presence of a third-order PT, $\partial_{E}^{2}S$ admits an inflection point whose numerical value remains constant across lattice sizes. As a result, the specific heat develops a peak that remains rounded and does not diverge in the thermodynamic limit. This mechanism is fundamentally different from the case of a second-order PT, where $\partial_{E}^{2}S$ exhibits a (negative) maximum that tends to zero with increasing system size, leading to the characteristic divergence of $C_v$. 

Importantly, these non-divergent, i.e., analytical thermodynamic observables are not artifacts of the microcanonical ensemble, nor are they finite-size effects. Rather, they are genuine manifestations of collective phenomena that do not necessarily produce catastrophic signatures. This interpretation is supported by the phenomenological analysis of the field histograms. Across the transition energy $\epsilon^{ind}_{3}$, the system shifts from an \emph{ordered} and \emph{hierarchical} phase to an \emph{almost-diffusive} phase. 

Below $\epsilon^{ind}_{3}$, the field variables are strongly peaked around $\phi=0$, with configurations $\phi=n\pi$ becoming increasingly less probable as $n$ increases; see Fig.~\ref{fig:histogram_phi_above_below_thresh}\textbf{a.1)}. Above $\epsilon^{ind}_{3}$, the system explores configurations with larger multiples of $\pi$ more freely; see Fig.~\ref{fig:histogram_phi_above_below_thresh}\textbf{b.1)}. In other words, states that were suppressed in the ordered phase become equally probable in the diffusive phase. This effect is clearly captured by the histogram of $\cos(\phi)$ shown in Fig.~\ref{fig:histogram_cosphi}. Once the energy threshold $\epsilon_{3}^{ind}$ is exceeded, the system occupies states associated with both positive and negative cosine values with comparable probability, producing two symmetric peaks. 

In conclusion, this work provides insights into the theoretical framework of phase transitions that call for a revision of their traditional definition. RG theory implicitly weakened the original ELY definition by focusing on universal scale-invariant structures, characterized by non-analytic behavior in correlation functions rather than in thermodynamic observables. However, the presence of a divergence in correlation functions does not necessarily imply a divergence in thermodynamic quantities. 

Our analysis of the 2D sine-Gordon model shows that the roughness PT, while accepted as genuine, manifests through persistent morphological features in entropy derivatives rather than divergences. This indicates that analyticity does not preclude the existence of a true PT, which deserves the same conceptual standing as transitions characterized by catastrophic singularities. This consideration is significant because dismissing non-divergent transitions as mere \emph{crossovers} risks overlooking genuine collective phenomena, as exemplified by the roughness PT itself. Moreover, such a reformulation does not replace existing theories but extends their framework to include higher-order transitions. 

Finally, this study opens a broader perspective on PTs, proposes a richer framework for understanding transitions in diverse physical and biological systems, and encourages further exploration of the role of higher-order PTs in nature.

\begin{acknowledgments}
This research was funded in part, by the Luxembourg National Research Fund (FNR), BroadApp C20/MS/14769845, and by European Research Council (ERC) Project FITMOL-101054629. For the purpose of open access, the authors have applied a Creative Commons Attribution 4.0 International (CC BY 4.0) license to any Author Accepted Manuscript version arising from this submission.\\
The authors thank Dr. Matteo Barborini for fruitful helps in coding, Dr. Dahvyd Wing and Dr. Igor Poltavsky for useful comments about the manuscript. The calculations presented in this paper were carried out using the HPC facilities of the University of Luxembourg~\cite{VBCG_HPCS14} {\small (see \href{http://hpc.uni.lu}{hpc.uni.lu})} and those of the Luxembourg national supercomputer MeluXina. 
\end{acknowledgments}

\appendix

\section{Configurational microcanonical Monte Carlo algorithm}
\label{app:micro_montecarlo}
The microcanonical ensemble has been reproduced by adopting the method proposed in Refs.~\cite{ray1996microcanonical,ray1991microcanonical} and consisting of a microcanonical Monte Carlo (MICROMC) algorithm in the configuration space. We have chosen a method based on a random proposal to enforce ergodicity. 

\subsection{Microcanonical distribution function}
The microcanonical algorithm is based on the identification of a suitable density probability function that is used to produce a reliable sampling. To find such a probability function, we start with the microcanonical partition function:
\[
\Omega(E) = \int \delta(H[\pi, \phi] - E) D\pi\,D\phi\,,
\]
where
\[
    D\pi=\prod_{\bm{n}\in\mathbb{L}}d\pi_{\bm{n}},\qquad D\phi=\prod_{\bm{n}\in\mathbb{L}}d\phi_{\bm{n}}\,,
\]
and where the Hamiltonian is
\[
    H(\pi, \phi) = \sum_{\bm{n}\in\mathbb{L}} \frac{\pi_{\bm{n}}^2}{2}  + V(\phi)\,,
\]
for any potential function. Finally, $\delta$ represents the Dirac delta function. From now on, we introduce
$M=N^2$ to denote the total number of degrees of freedom. We focus on the momentum integral that separates as

\begin{equation}\label{def:integral_momentum_micro}
    I_p = \int \delta \left(\sum_{\bm{n}\in\mathbb{L}} \frac{\pi_{\bm{n}}^2}{2} + V(\phi) - E\right)D\pi\,.
\end{equation}
Thus, we notice that the kinetic energy variable,
\[
    K = \sum_{\bm{n}\in\mathbb{L}} \frac{p_{\bm{n}}^2}{2}\;,
\]
can be interpreted as the equation for a $M$-dimensional sphere of radius $r^2=2K$. Then, exploiting the spherical coordinates in $M$ dimensions, we have
\[
    \prod_{\bm{n}\in\mathbb{L}}d\pi_{\bm{n}} = \rho^{M-1}\,d\rho\,d\Omega_{M-1}\,,
\]
where \( d\Omega_{M-1} \) is the differential solid angle element on the unit sphere \( S^{M-1} \), expressed as:
\[
    d\Omega_{M-1} = \prod_{i=1}^{M-1} d\theta_i \sin^{M-1-i} \theta_i,\qquad
    \Omega_{M}=\int d\Omega_{M-1}\;.
\]
Then, we define 
\[
    \rho=\sqrt{2K},\quad d\rho=\frac{dK}{\sqrt{2K}}\implies \rho^{M-1}d\rho=(2K)^{M/2-1}\,dK\,.
\]
Finally, integral \eqref{def:integral_momentum_micro} rewrites
\begin{equation}\label{}
\begin{split}
    I_p &= \Omega_{M}\int \delta \left[K-(E- V(\phi))\right](2K)^{M/2-1}\;dK\\
    &= 2^{M-1}\Omega_{M}(E- V(\phi))^{M/2-1}\Theta(E - V(\phi))\,.
\end{split}
\end{equation}
Reintroducing above the integral over field configurations, we recover the \emph{configurational microcanonical partition function} that reads
\begin{equation}\label{def:configurational_entropy}
    \Omega(E) = \int  (E - V(\phi))^{N/2 - 1} \Theta(E - V(\phi))\;D\phi.
\end{equation}
Notice that we have dropped irrelevant constants that play no role in the calculation of expectation values.\\

The numerical algorithm then exploits the probability distribution function arising from Eq.~\eqref{def:configurational_entropy}
$$
W_{E}[\phi]= \left(E-V(\phi)\right)^{M/2-1}\Theta[E-V(\phi)]\,.
$$
The sampling is carried out by randomly picking a lattice site, $\bm{n}$, and proposing a new random configuration $\phi^{old}_{\bm{n}}\mapsto \phi^{new}_{\bm{n}}=\phi^{old}_{\bm{n}} + \eta\Delta\phi$ where $\eta$ is a random number uniformly sampled from the interval $[-1,1]$ and $\Delta\phi$ is adjusted to achieve a target acceptance rate of $50\%-60\%$. The new configuration is accepted according to the Monte Carlo acceptance probability
$$
    W(\phi^{old}\to\phi^{new}) = \min\left(1,\frac{W_{E}[\phi^{new}]}{W_{E}[\phi^{old}]}\right)\,.
$$
Notice that an efficient way to evaluate this acceptance probability is to rewrite the acceptance ratio as follows:
$$
    \frac{W_{E}[\phi^{new}]}{W_{E}[\phi^{old}]} = \exp\left[\bigg(\frac{M}{2}-1\bigg)\log\left(\frac{E-V(\phi^{new})}{E-V(\phi^{old})}\right)\right]\,.
$$
once we have checked that $E-V(\phi^{new})>0$.

\subsection{Initial configuration} 

Initial conditions are randomly proposed in order to start with a high-energy configuration, $E_{rand}$, larger than the desired input energy $E_{inp}$ and such that $E_{rand}-V(\phi_{ini})>0$. Then, we perform $10^{4}$ steps of equilibration with the MICROMC algorithm. At this stage, we search for the correct value for $\Delta\phi$. To do that, we start with a small value for $\Delta\phi$, say $0.001$ and run a few MICROMC steps using the desired input energy $E_{inp}$ and computing the acceptance rate, $N_{\text{acc}}$. If $N_{\text{acc}}\notin[0.5,0.6]$, then we repeat the procedure by replacing $\Delta\phi$ with $\Delta\phi+0.001$. It should be stressed that other strategies, such as the molecular dynamics algorithm, have also been used to select the initial conditions. Comparison of the thermodynamic observables obtained with both methods did not yield appreciable differences. Finally, once the suitable tune parameter's value has been obtained, the configuration is equilibrated for $10^4$ steps, and finally the trajectory is evolved with the MICROMC method and used for computing averages. For each energy value, we have produced $N_{trj}=20$ realizations for the system's size $N=32,\,64$, while we produced $N_{trj}=80$ realizations for $N=128,\,256$. The specific energy $\varepsilon=E_{inp}/N^2$ is chosen within the interval $[0.350,3.525]$ for $N=32,\,64,\,128$ and $[0.5,1.5]$ for $N=256$. Finally, the sampling in energy is $\Delta \varepsilon=0.025$. Each thermodynamic observable has been evaluated through $N_{\text{avg}}=10^{6}$ measurements, performed in every $N_{\text{step}}=100$ MICROMC step. The microcanonical average of a given observable, $f$, is computed by
\begin{equation}\label{def:time_average}
    \langle f\rangle_{\varepsilon}=\frac{1}{ N_{\text{trj}}\cdot N_{\text{avg}}}\sum_{i=1}^{N_{trj}}\sum_{\alpha=1}^{N_{\text{avg}}}f(\phi^{(i)}_{\alpha})\,,
\end{equation}
where $f$ is the observable evaluated on the configuration $\phi^{(i)}_{\alpha}$ at the $\alpha$-th MC step for the $i$-th realization.

\subsection{Numerical calculation of microcanonical entropy's derivatives}
\label{app:PHT_method}
The MIPA method requires the estimation of higher-order derivatives of the microcanonical entropy from a microcanonical sampling of the phase space. To this purpose, we adopt the Pearson-Halicioglu-Tiller (PHT) method \cite{pearson1985laplace} which allows a direct calculation of the first- and second-order derivatives of the microcanonical entropy as follows. Any energy derivative of the entropy function can be rewritten in terms of averages of power of the kinetic energy $k=(E-V(\phi)/M$. In this framework, the first- and second-order derivatives of the microcanonical entropy can be written as 
\begin{equation}
\label{eq:I_II_derMicroCanS}
\begin{split}
     \partial_{\varepsilon}S(\varepsilon)= \left(\frac{1}{2}-\frac{1}{M}\right)&\langle k^{-1}\rangle_{\varepsilon}\,, \\
    \partial^{2}_{\varepsilon}S(\varepsilon)
    =M\bigg[\bigg(\dfrac{1}{2}-\dfrac{1}{M}\bigg) &\bigg(\dfrac{1}{2}-\dfrac{2}{M}\bigg)\langle k^{-2} \rangle_{\varepsilon}
    \\
    - &\bigg(\dfrac{1}{2}-\dfrac{1}{M}\bigg)^2 \langle k^{-1}\rangle^2_{\varepsilon} \bigg]\,,
\end{split}
\end{equation}
where the microcanonical averages have been estimated by Eq.~\eqref{def:time_average} over the realizations. Finally, the third-order derivative of entropy is computed using the numerical derivative of $\partial^{2}_{\epsilon}S(\epsilon)$, that is, by means of the finite difference (central) method.

\subsection{Histograms}

We construct a two-dimensional histogram that represents the frequency of field values across discrete intervals and bins. The procedure can be conceptually described as follows. \\
\textbf{Domain Division}. The entire range of possible field values is partitioned into a set of contiguous intervals of equal width. Each interval represents a subset of the domain, providing a coarse-grained categorization of the values. \\
\textbf{Subdivision of Intervals}. Each interval is further subdivided into a fixed number of bins, creating finer resolution within the interval. This hierarchical division allows for detailed tracking of variations within each interval. \\
\textbf{Field Value Assignment}. For each lattice site, the field value is evaluated, and its corresponding interval is determined on the basis of the position of the value within the global range. Within the selected interval, the precise bin is identified by determining the position of the value relative to the interval boundaries. \\
\textbf{Frequency Counting}. Once the interval and bin for a given field value are identified, the count for that bin is incremented. This process is repeated for all lattice sites, ensuring that the histogram accumulates the frequency distribution of field values across the entire lattice.\\
\textbf{Output}. The resulting histogram provides a two-dimensional representation of the field value distribution. The rows correspond to intervals, while the columns represent bins within each interval. The numerical values in the histogram represent the frequency of field values that fall within the corresponding bin.\\
This procedure has been applied for both $P[\phi]$ and $P[\cos(\phi)]$, whose main difference lies in the definition of the interval. For the histogram $P[\phi]$, we have chosen $[\phi_{min},\phi_{max}]$ with $\phi_{min}=-5000$, $\phi_{max}=|\phi_{min}|$ while for $P[\cos(\phi)]$ the interval is $[-1,1]$.\\
At any energy, we have built the histograms associated to each realization and then the representative histogram plots in the main text is obtained by averaging them over the total number of realization $N_{trj}$.

\section{Microcanonical analysis at fixed system size}
\label{app:mipa_fixed_size}
\begin{figure*}[tbh!]
    \includegraphics[height=8cm, width=17.5cm,keepaspectratio]{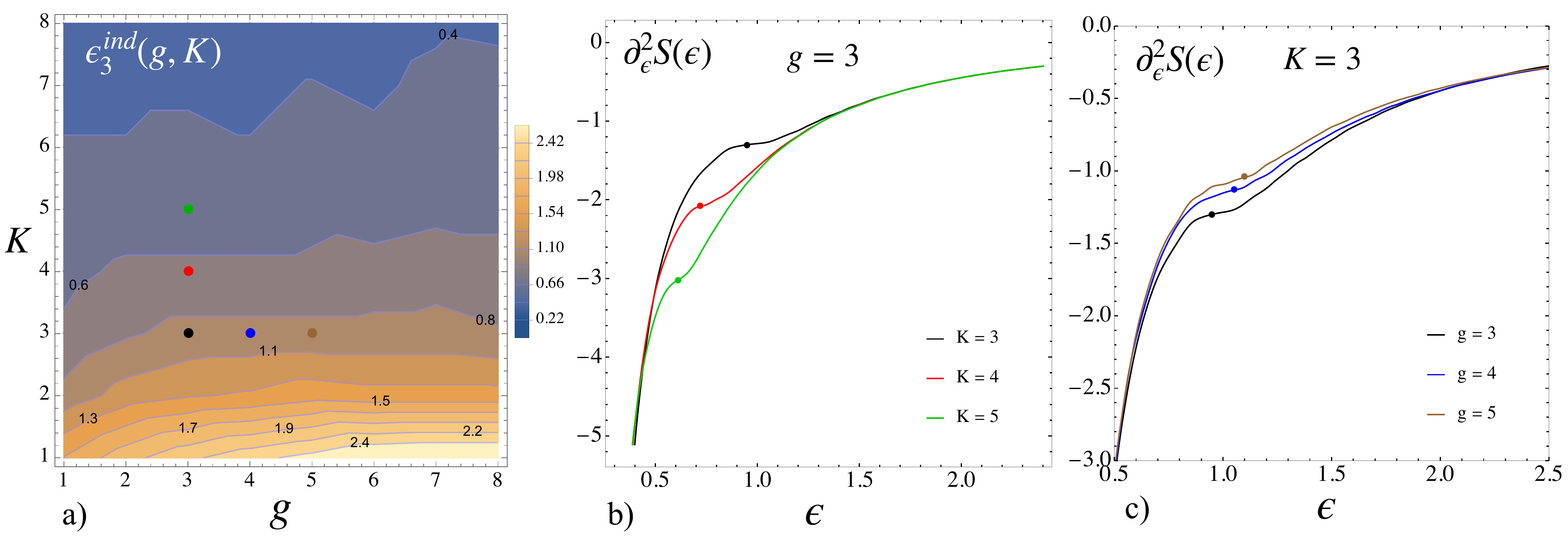} 
    \caption{\textbf{Microcanonical analysis and critical energy landscape for the third-order PT at lattice size $N=32$}. In plot \textbf{a)}, we report the energy values where the third-order PT is detected by MIPA for any values of the system's parameters $(g,K)$ studied in this work: $g=\{1,8\}$ and $K=\{1,8\}$. In plot \textbf{b)}, the second-order derivative of entropy, $\partial^{2}_{\epsilon}S(\epsilon)$, is reported for $g=3$ and $K=3,\,4,\,5$. Similarly, in plot \textbf{c)}, $\partial^{2}_{\epsilon}S(\epsilon)$, is reported for $K=3$ and $g=3,\,4,\,5$. These curves are obtained through interpolation of the numerical data. The colored points on the critical energy landscape, plot \textbf{a}), indicate the values of the pair $(g,K)$. These colors are also adopted for the curves in plots \textbf{b)} and \textbf{c)}. Here, the colored points locate the inflection points of $\partial^{2}_{\epsilon}S(\epsilon)$ for a given pair of parameters values that are reported in the legend. In so doing, we can easily detect the energy values of the third-order PT. The black point represents the case investigated in the thermodynamic analysis, namely, $g=3$ and $K=3$. In plot \textbf{a}), the quasi-horizontal lines crossing the diagram represent the borders of regions with different transition energy values. Small variations in the parameter $K$ lead to larger variations in the transition energy and viceversa for the parameter $g$.}
      \label{fig:phase_diagram_third_order_PT}  
\end{figure*}

We apply MIPA to the 2D-sG model with fixed system size, i.e. $N=32$, on a grid of parameter values $(g,K)$ with $g=\{1,8\}$ and $K=\{1,8\}$. Thus, we produce $8^2$ sets of thermodynamic observables, namely $\{\partial_{\epsilon}S(\epsilon),\partial^{2}_{\epsilon}S(\epsilon),\partial^{3}_{\epsilon}S(\epsilon)\}_{g,K}$. Microcanonical analysis has detected the existence of a third-order PT at any chosen parameter values. In Fig.~\ref{fig:phase_diagram_third_order_PT} plot \textbf{a)}), we report the critical energy landscape for the third-order PT at lattice size $N=32$. This is a contour plot where the energy values $\epsilon_{3}^{ind}$, associated with the third-order PT detected by MIPA, are plotted versus the pair $(g,K)$. Then, we have chosen five sets of parameters $(g,K)$ and shown $\partial^2_{\epsilon}S$ in plots ~\ref{fig:phase_diagram_third_order_PT}\textbf{b)} and \ref{fig:phase_diagram_third_order_PT}\textbf{c)}. In these plots, we observe a visible inflection point at every pair $(g,K)$ which is marked by colored points. In plot~\ref{fig:phase_diagram_third_order_PT}\textbf{b)}, we plot the second-order derivative at fixed $g=3$ while varying $K$, while, in plot~\ref{fig:phase_diagram_third_order_PT}\textbf{c)}, we fix $K=3$ and vary $g$. Interestingly, in the critical energy landscape (plot~\ref{fig:phase_diagram_third_order_PT}\textbf{a)}), we observe several quasi-horizontal lines corresponding to quasi-fixed values of $K$. This indicates that the critical energy values do not change appreciably on a wide range of $g$'s values when $K$ is fixed and it can be easily seen in plot~\ref{fig:phase_diagram_third_order_PT}\textbf{c)}, where the inflection points of $\partial_{\epsilon}S(\epsilon)$ do not move as much as for the case with fixed $g$ and varying $K$, see plot~\ref{fig:phase_diagram_third_order_PT}\textbf{b)}.

\newpage



~
\clearpage
\bibliography{biblio}

\end{document}